\documentclass[showpacs,superscriptaddress, twocolumn, pra]{revtex4}
\usepackage{amsmath}
\usepackage{amsfonts}
\usepackage{amssymb}
\usepackage{graphicx}

\begin{document}

\title{Detection of $N$-particle entanglement with generalized Bell inequalities}

\author{Wies{\l}aw Laskowski}
\affiliation{Instytut Fizyki Teoretycznej i Astrofizyki
Uniwersytet Gda\'nski, PL-80-952 Gda\'nsk, Poland}
\author{Marek {\. Z}ukowski}
\affiliation{Instytut Fizyki Teoretycznej i Astrofizyki
Uniwersytet Gda\'nski, PL-80-952 Gda\'nsk, Poland}
\affiliation{Institut f\"ur Experimentalphysik, Universit\"at
Wien, Boltzmanngasse 5, A--1090 Wien, Austria}
\affiliation{Tsinghua University, Beijing, China}
\date{\today}

\begin{abstract}
We show that the generalized Bell-type inequality, explicitly involving rotational symmetry of physical laws, is very efficient in  distinguishing between true $N$ particle quantum 
correlations and correlations involving less particles.  This applies to various types of generalized partial separabilities.
We also give a rigorous proof that the new Bell inequalities are maximally violated by the GHZ states, and find a very handy description of the $N$-qubit correlation function.
\end{abstract}


\maketitle
\section{Introduction}
The correlations of quantum systems containing more then two 
particles are an extensive field of theoretical and experimental study. Before the trailblazing paper of Greenberger, Horne and Zeilinger \cite{GHZ} in 1986 Svetlichny 
derived a Bell-type inequality \cite{SVETLICHNY}, violated by quantum 
mechanical predictions for pure three-qubit states, with the 
property that it can distinguish between three-qubit 
correlations and two-qubit ones. 
A generalization of the results by Svetlichny was given in e.g. \cite{SEEVINCK, COLLINS, TOTH}. 
In \cite{COLLINS} one can find an inequality for $N$ qubits, which can be used to  classify
the $N$-body correlations in the following way. If any quantum state 
violates the inequality by a factor larger than ${2}^{\frac{1}{2}(N-2)}$, then 
the state contains true $N$-body correlations. The factor that distinguishes maximal violation of the inequality, which occurs for a GHZ state, and the maximal possible violation by a state with one qubit completely unentangled with other $N-1$ ones is $\sqrt{2}$. Interestingly in Ref. \cite{TOTH} it was shown that a suitable choice of measurement settings allows to increase this factor, for three qubits, up to two (this is important, because the violation factor is directly linked with the resistance, with respect to noise admixtures, of the non-classical correlations).  Here we investigate the Bell inequalities of a type introduced in \cite{ZUKOWSKI1993, ROTATION}, which include the assumption that physical laws must have a rotationally invariant form. They are dependent on the measured values of the correlation tensor for the $N$-qubit processes under investigation. What is to be measured are $2^N$ components of the tensor, thus the same number of measurement settings is involved as for the standard Bell inequalities.  The factor of violation of such an inequality for a GHZ state is two times higher than for any biseparable state.   The results are very easily generalizable to more complicated situations. We also give a rigorous proof that the Bell inequalities are maximally violated by the GHZ states, and find a very handy description of the $N$-qubit correlation function.

 As it was shown in Ref. \cite{ROTATION} the generalized Bell inequalities can be utilized to reveal to what extent realistic models can reproduce  correlation functions $E(\vec{a}_1,..., \vec{a}_N)$, which have a rotationally invariant form. It is assumed, that correlation functions are  linear in vectors  $\vec{a}_i$, which specify the dichotomic local qubit observables (i.e., $\vec{a}_i \cdot \vec{\sigma}$, where $\vec{\sigma}$ stands for the three Pauli operators). Note that such is the case for the quantum correlation functions.  

Let us present the basic elements of our reasoning. On one hand, in the case of local and realistic theories, the correlation 
function for the qubits must be expressible in the following form:
\begin{equation}
E_{HV} (\vec a_1(\alpha_1), ...,\vec a_N(\alpha_N)) = \int \textrm{d} \lambda \rho(\lambda)
\prod_{i=1}^N I_i(\alpha_i),
\end{equation}
where $\rho(\lambda)$ is a certain distribution function of some (hidden) parameters $\lambda$, 
and $I_i$ is a function, that predetermines the 
values of results of experiments that can be performed on the given local system.
Its allowed values are  the eigenvalues of the local  dichotomic observables, $\pm 1$.
Finally $\alpha_i$ is a certain parametrization of the local setting $\vec a_{i}$. It is introduced to facilitate further manipulations.

On the other hand, the $N$-qubit correlation functions, in quantum theory, 
have the following form:
\begin{equation}
E(\vec{a}_1,...,\vec{a}_N) = \hat T \bullet ( \vec a_1 \otimes ... \otimes \vec a_N),
\label{QM-CORRELATION}
\end{equation}
where $\hat T$ is the correlation tensor for a quantum state, $\rho$, 
with components given by 
\begin{equation}
T_{i_1...i_N}=E(\vec{x}_{i_1},...,\vec{x}_{i_N})
\end{equation}
with $\vec{x}_{i_k}$, $i_k=1,2,3$ representing some three (local) basis vectors for the $k$th observer. By
the symbol $\bullet$ we represent the scalar product in $R^{3N}$. 
Note that the left hand side of Eq. (\ref{QM-CORRELATION}) gives the general form of the class of  correlation functions of a rotationally invariant form 
mentioned in the introduction. 
If one constrains the 
measurement settings of each observer to just one plane (which for each of them 
can be different) the measurement direction vectors 
can be expressed by $\vec a_i(\alpha_i) = \cos{\alpha_i} \hat{y}_i + \sin{\alpha_i} \hat{x}_i$, 
where $\hat{x}_i, \hat{y}_i$ are two basis vectors of $R^3$ (which can be individually defined by each observer). In such a case, 
the  correlation function is a scalar given by: 
\begin{eqnarray}
E(\alpha_1, ..., \alpha_N) &=& \sum_{i_1...i_N = 1,2} T_{i_1...i_N} \sin{(\alpha_1 +  (i_1-1) \frac{\pi}{2})} \nonumber \\ &\times&  ... \sin{(\alpha_N +  (i_N-1) \frac{\pi}{2})}. 
\label{EXPANSION} 
\end{eqnarray}

\section{The Bell inequality}

In order to show contrast between quantum and local realistic predictions 
for correlation function we use a simple geometrical fact: 
{\em Two vectors $\vec h$ and $\vec q$ are equal if, and only if, their 
scalar product $(\vec h,\vec q)$ is equal to the square of the norms 
of the two vectors: $(\vec h,\vec q) = ||\vec h||^2 = ||\vec q||^2$. 
Thus, $(\vec h,\vec q) \leq ||\vec q||^2$ implies that $\vec h \neq \vec q$}. 
In our case the vectors are replaced by correlation functions, and we use the scalar product
of a real Hilbert space of square integrable functions of $\alpha_i$, $i=1,..., N$, that is 
\begin{equation}
(h,q)=\int_0^{2\pi} \textrm{d} \alpha_1 ... \int_0^{2\pi} \textrm{d} \alpha_N h(\alpha_1,\ldots,\alpha_N)q(\alpha_1,\ldots,\alpha_N). \label{SCALAR}
\end{equation}

Let us calculate an upper bound of the scalar product of $E_{HV}$ and $E$. We get:
\begin{eqnarray}
(E_{HV}, E) &\leq& \int_0^{2\pi} \textrm{d} \alpha_1 ... \int_0^{2\pi} \textrm{d} \alpha_N
\prod_{l=1}^N I(\alpha_l) \nonumber \\ &\times& 
\sum_{i_1, ..., i_N=1,2} T_{i_1...i_N} \sin{(\alpha_1 +  
(i_1-1) \frac{\pi}{2})} ... \nonumber \\
&\times& \sin{(\alpha_N +  (i_N-1) \frac{\pi}{2})} \label{semi}\\
&=& \sum_{i_1, ..., i_N=1,2} T_{i_1...i_N} \nonumber \\
&\times& \prod_{l=1}^N \int_0^{2\pi} 
\textrm{d} \alpha_l I(\alpha_l) \sin{(\alpha_l +  (i_l-1) \frac{\pi}{2})}. \nonumber
\label{fin}
\end{eqnarray}
The sum (\ref{semi}) has $2^N$ elements containing all 
possible combinations of sine and cosine functions. 
In Ref. \cite{ROTATION} it was shown that 
\begin{equation}
(E_{HV}, E) \leq 4^N E_{{max}},
\label{ineq}
\end{equation}
where $E_{max}$ is the maximal possible value of a correlation function, that is $E_{max} = \max_{\alpha_1, \ldots, \alpha_N} E(\alpha_1,\ldots, \alpha_N)$. It is important to stress that the maximalization is only within some fixed planes of observation defined for each observer by his/her local versors $\hat{x}_i$ and $\hat{y}_i$. 
Note that this inequality depends on the correlation function for quantum state to be analyzed. Interestingly, one can use also any correlation function that has the rotationally invariant form, as the one given in Eq. (\ref{QM-CORRELATION}). This generalized Bell inequality, Ineq. (\ref{ineq}), is the basis of our further considerations.  

If one replaces in the above inequality $E_{HV}$ by $E$ one gets:
\begin{eqnarray}
(E,E) &=& ||E||^2 \nonumber \\
&=& \int_0^{2\pi} \textrm{d} \alpha_1 ... \int_0^{2\pi} \textrm{d} \alpha_N
(T_{1...1} \sin{\alpha_1} ... \sin{\alpha_N} \nonumber \\
&+& ... +  T_{2...2} \cos{\alpha_1} ... \cos{\alpha_N})^2. 
\label{ub1}
\end{eqnarray}
Since $\int_0^{2\pi}\textrm{d}\alpha \sin{\alpha} = 0$ 
and $\int_0^{2\pi} \textrm{d}\alpha \sin^2{\alpha} = \pi$
we have $(E, E) = \pi^N \sum_{i_1,...,i_N=1,2} T_{i_1...i_N}^2$. This expression differs very much from the right hand side of Ineq. (\ref{ineq}), 
and thus for many quantum states one can find strong violations of this bound. 
As a matter of fact these violations exceed, in the case of {\em four} or more qubit GHZ states, 
those that can be obtained with any standard Bell inequality (i.e., a two-setting per observer one), see \cite{ROTATION}.

If  
\begin{equation}r=4^{-N} \frac{||E||^2}{E_{max}}>1 \label{R}\end{equation} 
the new Bell inequalities are violated. That is a correlation function $E$ for which Ineq. (\ref{R}) holds cannot have a local realistic model. The factor $r$ is a good measure of noise 
robustness of the non-classical correlations in a given quantum state $\rho_o$. 
Take a  state, given by $$\rho(V)=V\rho_o +(1-V)\rho_{noise},$$
where $\rho_{noise}= \openone /d$, where 
in turn $\openone$ is the unit operator, and $d=2^N$ stands for the dimension 
of the Hilbert space. Obviously, $0 \leq  V \leq 1$. Such a  state 
violates the given Bell inequality by if and only if $V>1/r$.

\section{Partial separability}

Let us discuss how does the inequality fare in the context of discretion between partially separable states and states that have true $N$ 
qubit entanglement. We start first with the case of biseparable states, for which the reasoning will be presented in some detail. Next we discuss 
generalizations. 

\subsection{Biseparability}

Our definition of biseparability runs as follows. As biseparable we shall consider all states, $\rho_{1...N}^{bisep}$, for which there exist a convex expansion into density matrices, $\rho_I$, where $I$ is some index, such that every density matrix $\rho_I$  factorizes into at least two density matrices of two separate subsystems, one consisting of $N-k$ qubits, and the other of $K$ qubits, i.e. $\rho_I=\rho_{A_I}\otimes\rho_{B_I}$ (subsystems $A_I$ and $B_I$ may vary, depending on $I$).   
 For example for $N=4$ we can have the decomposition in which the following classes of partially factorable density matrices may pop up $\rho_i \otimes \rho_{jkl}$ and $\rho_{ij} \otimes \rho_{kl}$, where $i,j,k,l=1,...,4$ enumerate the qubits. Note that density operators of the form  $\rho_1 \otimes \rho_{2}\otimes \rho_{3} \otimes \rho_{4}$ are joint members of the two classes.

By $r_{bisep}$ let us denote  the violation 
factor of (\ref{ineq}) by a biseparable state. 
Then, {\em if the the $\rho_{1...N}$ state violates 
the inequality (\ref{ineq}) by a factor $r$ such that 
\begin{equation}
 r>r^{MAX}_{bisep},
\label{NB}
\end{equation}
where $r^{MAX}_{bisep}$ is the maximal value of $r$ for any biseparable state,
then $\rho_{1...N}$ must contain the true non-classical $N$-particle correlations, i.e. cannot be biseparable}.

We shall calculate $r^{MAX}_{bisep}$, but 
first we show that $$r_{1...N} \equiv \max_{\rho}\left(||E||^2 /4^N E_{max}\right),$$
that is the maximal possible violation factor of inequality by any state $\rho$, is equal to $\frac{1}{2}( \pi/2)^N$. 

Let us start from the case of two qubits. Our first task is to find the general form the two-qubit correlation function, for measurements  in $x_i-y_i$ planes, $i=1,2$.  The correlation function is defined as
\begin{equation}
E(\alpha_1, \alpha_2) = \textrm{Tr}[\rho   \sigma^{(1)}(\alpha_1) \otimes \sigma^{(2)}(\alpha_2)],
\end{equation}
where $\sigma^{(i)}(\alpha_i) = \cos{\alpha_i}\sigma^{(i)}_x + \sin{\alpha_i}\sigma^{(i)}_y $. Note that for the operator $\sigma^{(1)} \otimes \sigma^{(2)}$ expressed 
 in the $z_1$-$z_2$ product basis. 
 only the antidiagonal elements do not vanish. If one denotes the element of the basis by 
 $\{ |00\rangle, |01\rangle, |10\rangle, |11\rangle \}$,
 the non-zero elements are given by $\{\sigma^{(1)} \otimes \sigma^{(2)}\}_{k,l;1-k,1-l} = \exp(-i(-1)^l (\alpha_1 + (-1)^{l-k} \alpha_2))$. Therefore,
\begin{eqnarray}
E(\alpha_1, \alpha_2) &=&  2 (\cos{(\alpha_1 + \alpha_2)} ~ \textrm{Re} \rho_{0,0;1,1} \nonumber \\
&+&\cos{(\alpha_1 - \alpha_2)} ~ \textrm{Re} \rho_{0,1;1,0}\nonumber \\
&-&\sin{(\alpha_1 + \alpha_2)} ~ \textrm{Im} \rho_{0,0;1,1} \label{e} \\
&-&\sin{(\alpha_1 - \alpha_2)} ~ \textrm{Im} \rho_{0,1;1,0}).\nonumber 
\end{eqnarray}
Using Schwarz inequality we can show that the function (\ref{e}) is bounded from above by $2 (|\rho_{0,0;1,1}| + |\rho_{0,1;1,0}|)$. Please note that this value is actually achieved by E for $\alpha_1 = (\Phi_{0,0;1,1} + \Phi_{0,1;1,0})/2$ and $\alpha_2 = (\Phi_{0,0;1,1} - \Phi_{0,1;1,0})/2$, where $\Phi_{k,l;1-k,1-l}$ denotes the argument of the complex number $\rho_{k,l;1-k,1-l}$. Therefore $\max_{\alpha_1,\alpha_2} E(\alpha_1,\alpha_2) = 2 (|\rho_{0,0;1,1}| + |\rho_{0,1;1,0}|)$.
This result can be easily generalized for an arbitrary number of
parties. For $N$ qubits the general form of correlation function for measurements restricted to  $x_i-y_i$ planes reads:
\begin{widetext}
\begin{eqnarray}
E(\alpha_1, ..., \alpha_N)= 2\sum_{i_2,...,i_{N-1} = 0,1} \big( \cos{(\alpha_1 + (-1)^{i_2} \alpha_2 + ... + (-1)^{i_N} \alpha_N)} ~\textrm{Re} ~ \rho_{0,i_2,...,i_N;1,1-i_2,...,1-i_N} \nonumber \\
- \sin{(\alpha_1 + (-1)^{i_2} \alpha_2 + ... + (-1)^{i_N} \alpha_N)} ~\textrm{Im} ~ \rho_{0,i_2,...,i_N;1,1-i_2,...,1-i_N} \big)
\label{CORR-DENSITY}
\end{eqnarray}
\end{widetext}
and its maximal value is given by:
\begin{eqnarray}
E_{max} &=& \max_{\alpha_1,...,\alpha_N} E(\alpha_1,...,\alpha_N) \nonumber \\
&=& 2 \sum_{k_2,...,k_N=0,1} |\rho_{0,k_2,...,k_N;1,1-k_2,...,1-k_N}|.
\end{eqnarray}

Let us express the value of $(E,E)=||E||^2$ in terms of the parameters of the density matrix. We simply put the expression given in Eq. (\ref{CORR-DENSITY})
into the integral form of $||E||^2$. Since all cosine and sine functions, which appear in (\ref{CORR-DENSITY}), are orthogonal to each other (with respect to the scalar product defined by (\ref{SCALAR})), and all these functions have the same norm, equal $\frac{1}{2}(2\pi)^N$, the final result reads  
\begin{equation}
||E||^2 = 2(2\pi)^N \sum_{k_2,...,k_N=0,1} |\rho_{0,k_2,...,k_N;1,1-k_2,...,1-k_N}|^2.
\end{equation}

Let us calculate the maximum of $r_{1...N}$.
We get 
\begin{widetext}
\begin{eqnarray}
r_{1...N} &=& \frac{||E||^2}{4^N E_{max}} \nonumber  \\ 
&=& \frac{2(2\pi)^N  \sum_{k_2,...,k_N=0,1} |\rho_{0,k_2,...,k_N;1,1-k_2,...,1-k_N}|^2}{4^N   2 \sum_{k_2,...,k_N=0,1} |\rho_{0,k_2,...,k_N;1,1-k_2,...,1-k_N}|} \nonumber \\
&\leq& \frac{2(2\pi)^N \sum_{k_2,...,k_N=0,1} |\rho_{0,k_2,...,k_N;1,1-k_2,...,1-k_N}|^2}{4^N   4 \sum_{k_2,...,k_N} |\rho_{0,k_2,...,k_N;1,1-k_2,...,1-k_N=0,1}|^2} \label{im} \\
 &=&  \frac{1}{2}( \pi/2)^N \nonumber
\end{eqnarray}
\end{widetext}
The  inequality in  (\ref{im}) is true because the module of any antidiagonal elements of the density matrix satisfy $|\rho_{antidiagonal}| \leq 1/2$. This can be easily checked for pure states, which, of course, give the upper bound. Thus the relation $2|\rho_{antidiagonal}|^2 \leq  |\rho_{antidiagonal}|$ used in the inequality is justified.
The obtained bound is actually achieved by $N$ qubit GHZ states. {Simply, the correlation tensor $\hat T$ of a $N$ qubit GHZ 
state, $\frac{1}{\sqrt{2}}(|+,\ldots,+>+|-,\ldots,->)$, where $|\pm>$ are eigenstates of $\sigma_z$,  for  observation directions given for the $i$-th observer by $\hat{x}_i$ and $\hat{y}_i$, has $2^{N-1}$ nonzero components equal
$\pm1$'s.  Therefore $||E||^2 / 4^N E_{max} = \frac{1}{2}( \pi/2)^N$.

 The maximum  $r_{bisep}^{MAX}$
can be calculated in the same way as above. To this end, note that the modulus of any antidiagonal element $|\rho_{antidiagonal}^{bisep}|$ of the biseparable density matrix is at most equal to 1/4.  This again can be checked by considering pure biseparable states. Thus, now  we have $4|\rho_{antidiagonal}|^2 \leq  |\rho_{antidiagonal}|$.
Using this fact in (\ref{im}) we get 
\begin{eqnarray}
r_{bisep}^{MAX} = \frac{||E^{(bisep)}||^2}{4^N E^{(bisep)}_{max}}  
= \frac{1}{4}( \pi/2)^N . 
\end{eqnarray}
Finally, the condition (\ref{NB}) one can rewritten in the form
\begin{equation}
\frac{1}{4}( \pi/2)^N   < r \leq \frac{1}{2}( \pi/2)^N .
\end{equation}
Please note that value of upper bound is two times higher than the lower one. 
That is, the inequality (\ref{ineq}), which we use for the construction 
of the condition (\ref{NB}), is more sensitive than the 
ones proposed in \cite{COLLINS}. The same sensitivity factor 
was obtained in \cite{TOTH}, however, it requires
a specific choice of observables. Interestingly, for $N=2$ and $N=3$ any violation 
of the inequality (\ref{ineq}) implies that the state has the true $N$-body correlations. 
As a matter of fact,  the state has true $N$-qubit correlation whenever $r$ 
of (\ref{ineq}) is greater than 0.62 and 0.97 for two and three qubit states, respectively.
That is, the inequality does not have to be violated.

\subsection{Other partial separabilities}

The presented method can be simply generalized to the case of $k$-separable states ($k>2$). 
We shall use the following definition of $k$-separable density matrices, $\rho_{1...N}^{k-sep}$. There must  exist for it  a convex expansion into density matrices, $\rho_I$,  where $I$ is some index, that is $\rho_{1...N}^{k-sep}=\sum_{I}P_I \rho_I$, with  $P_I\geq0$ and  $\sum_{I}P_I$,  such that every density matrix $\rho_I$  factorizes into at least $k$ density matrices of $k$ separate subsystems $S^I_n$, $n=1,\ldots, k$ (such subsystems can consist out of of minimum one up to $N-k +1$ qubits). That is,  $\rho_I=\otimes_{n=1}^{k}\rho^I_{S^I_n}$. Of course, the  splitting  into subsystems may vary, depending on $I$).   
Obviously, for any antidiagonal element of $k$-separable matrices
we have $|\rho_{antidiagonal}^{k-sep}| \leq (1/2)^k$. Therefore
\begin{eqnarray}
r_{k-sep}^{MAX}  =  2^{-k} (\pi/2)^N.
\end{eqnarray}
For fully-separable states, i.e. $k=N$, one has $r_{fully-sep}^{MAX} = (\pi/4)^N<1$. Again the inequality  (\ref{ineq}) does not  have to be violated to show that the state is not fully-separable.
Please note that for all $N$ and $k$ the threshold violation factor $r_{k-sep}^{MAX}$ is two times bigger than
$r_{(k-1)-sep}^{MAX}$. 

\section{Discussion}
As we have seen the approach is pretty efficient in discovering  quantum correlations of truly $N$ particle nature. 
What is important, to get the data needed to find the $r_{exp}$, all one needs to measure  are the values of the correlation functions at two settings per each observer, exactly like in the case of the standard Bell inequalities. The optimal choice is that the two settings of each observer  define complementary measurements. Having these it is a straightforward exercise to find $E_{max}$. One simply uses the formula (\ref{EXPANSION}) with
$T_{i_1...i_N}=E(\vec{x}_{i_1},...,\vec{x}_{i_N})_{exp}$, and searches for its maximal value. Please note, that exactly this method is implied by the derivation of the bound in (\ref{ineq}). Behind all this is the assumption of rotationally invariant form of the correlation function, which is well justified by the Noether Theorem, and its implications.

The other feature of the present approach is that it is very easy to formulate the general condition for an arbitrary number of qubits, $N$. Also it can be generalized to different type of partial separability than just the one involving two subsystems (bi-separability).

Let us finish with a purely speculative conjecture. It might be the case, that a conjunction of rotational invariance principle, which must hold for the form of physical laws, of local realism, and of some other principle, perhaps the one introduced by Zeilinger and Brukner \cite{ZEILINGER-BRUKNER}, may lead to an ultimate version of Bell's theorem for qubit systems. One can put forward this speculation, because  more generalized inequalities involving rotational invariance, of the type introduced in \cite{KASZL-ZUK}, lead to even more sensitive conditions for true non-classical $N$ qubit correlations than (\ref{ineq}), which in case of some specific examples are equivalent to non-separability criteria \cite{LASK-ZUK}.

The work is part of the MNiI Grant no. 1 P03 04927. MZ acknowledges Professorial Subsidy of FNP.
WL acknowledges a FNP stipend.

\end{document}